\documentclass[12pt,a4paper]{article}
\usepackage{amsmath}
\usepackage{latexsym}
\makeatletter
\def\rddots{\mathinner{\mkern1mu\raise\p@%
    \vbox{\kern7\p@\hbox{.}}\mkern2mu%
    \raise4\p@\hbox{.}\mkern2mu\raise7\p@\hbox{.}\mkern1mu}}
\makeatother
\newcommand{\ket}[1]{{\vert{#1}\rangle}}
\newcommand{\bra}[1]{{\langle{#1}\vert}}
\newcommand{\kett}[1]{{\vert{#1}\rangle\rangle}}
\newcommand{\braa}[1]{{\langle\langle{#1}\vert}}

\newcommand{\fukuso}{{\mathbf C}}
\newcommand{\real}{{\mathbf R}}

\newcommand{\tr}{{\rm tr}}

\begin{document}

\title{\sl Decoherence and Copenhagen Interpretation :\\ A Scenario}
\author{
  Kazuyuki FUJII
  \thanks{E-mail address : fujii@yokohama-cu.ac.jp }\\
  International College of Arts and Sciences\\
  Yokohama City University\\
  Yokohama, 236--0027\\
  Japan
  }
\date{}
\maketitle
\begin{abstract}
  In this paper we give a reasonable explanation (not 
proof) to the Copenhagen interpretation of Quantum 
Mechanics from the view point of decoherence theory. 
  
  Mathematical physicists with strong mission must prove 
{\bf the Copenhagen interpretation} at all costs.
\end{abstract}
\vspace{5mm}\noindent

\vspace{5mm}\noindent

\section{Introduction}
  When we start studying Quantum Mechanics the most 
difficult part to understand is the so--called Copenhagen 
interpretation. Usually beginners skip over this part, 
which is a wise choice in a certain sense. However, 
some researchers feel guilty about skipping over this.

  In this paper we try to give a proof to it from the view point 
of decoherence theory. Namely, we embed it into the theory 
of decoherence and solve a master equation based on 
density matrix (not wave function) exactly. 

We will perform this by both incorpolating the results in \cite{KF2}, 
\cite{KF3} and making the idea in \cite{KF1} clearer. 
The method is of course not complete, but some researchers 
may feel relieved. To the best of our knowledge this is the finest 
method up to the present.

\section{Principles of Quantum Mechanics}
In order to set the stage and to introduce proper notation, 
let us start with a system of principles of Quantum 
Mechanics (QM in the following for simplicity). See for example 
\cite{PD}, \cite{HG}, \cite{AP} and \cite{AH}. That is,

\vspace{5mm}\noindent
\begin{Large}
{\bf System of Principles of QM}
\end{Large}

\vspace{3mm}\noindent
1.\ {\bf Superposition Principle}\\
If $\ket{a}$ and $\ket{b}$ are physical states then their 
superposition $\alpha\ket{a}+\beta\ket{b}$ is also 
a physical state where $\alpha$ and $\beta$ are 
complex numbers.

\vspace{3mm}\noindent
2.\ {\bf Schr\"{o}dinger Equation and Evolution}\\
Time evolution of a physical state proceeds like
\[
\ket{\Psi}\ \longrightarrow\ U(t)\ket{\Psi}
\]
where $U(t)$ is the unitary evolution operator 
($U^{\dagger}(t)U(t)=U(t)U^{\dagger}(t)={\bf 1}$ and $U(0)={\bf 1}$) 
determined by a Schr\"{o}dinger Equation.

\vspace{3mm}\noindent
3.\ {\bf Copenhagen Interpretation}\footnote{There are some 
researchers who are against this terminology, see for 
example \cite{AH}. However, I don't agree with them 
because the terminology is nowadays very popular in the 
world}\\
Let $a$ and $b$ be the eigenvalues of an observable $Q$, and 
$\ket{a}$ and $\ket{b}$ be the normalized eigenstates corresponding to 
$a$ and $b$. When a state is a superposition $\alpha\ket{a}+\beta\ket{b}$ 
and we observe the observable $Q$ the state collapses like
\[
\alpha\ket{a}+\beta\ket{b}\ \longrightarrow \ \ket{a}
\quad\mbox{or}\quad
\alpha\ket{a}+\beta\ket{b}\ \longrightarrow \ \ket{b}
\]
where their collapsing probabilities are $|\alpha|^{2}$ and 
$|\beta|^{2}$ respectively ($|\alpha|^{2}+|\beta|^{2}=1$).

This is called the collapse of the wave function and 
the probabilistic interpretation.

\vspace{3mm}\noindent
4.\ {\bf Many Particle State and Tensor Product}\\
A multiparticle state can be constructed by the superposition of 
the Knonecker products of one particle states, which are called 
the tensor products. For example, 
\[
\alpha|{a}\rangle\otimes|{a}\rangle +\beta|{b}\rangle\otimes|{b}\rangle
\equiv 
\alpha|{a,a}\rangle +\beta|{b,b}\rangle
\]
is a two particle state.

\vspace{5mm}
The target of this paper is to give a proof to the Copenhagen 
interpretation, so we give a symbolic figure of it for the latter 
convenience (we take  $\ket{0}$ and $\ket{1}$ in place of 
$\ket{a}$ and $\ket{b}$ in the following).

\vspace{10mm}
\begin{center}
%WinTpicVersion3.08
\unitlength 0.1in
\begin{picture}( 57.4000, 48.1500)( 12.0000,-52.1500)
% CIRCLE 2 2 3 0
% 4 2195 1395 2235 1455 2255 1425 2255 1425
% 
\special{pn 8}%
\special{ar 2196 1396 72 72  0.0000000 0.1666667}%
\special{ar 2196 1396 72 72  0.6666667 0.8333333}%
\special{ar 2196 1396 72 72  1.3333333 1.5000000}%
\special{ar 2196 1396 72 72  2.0000000 2.1666667}%
\special{ar 2196 1396 72 72  2.6666667 2.8333333}%
\special{ar 2196 1396 72 72  3.3333333 3.5000000}%
\special{ar 2196 1396 72 72  4.0000000 4.1666667}%
\special{ar 2196 1396 72 72  4.6666667 4.8333333}%
\special{ar 2196 1396 72 72  5.3333333 5.5000000}%
\special{ar 2196 1396 72 72  6.0000000 6.1666667}%
% CIRCLE 2 0 3 0
% 4 3795 1395 3835 1455 3855 1425 3855 1425
% 
\special{pn 8}%
\special{ar 3796 1396 72 72  0.0000000 6.2831853}%
% STR 2 0 3 0
% 3 2215 895 2215 995 5 0
% $|0\rangle$
\put(22.1500,-9.9500){\makebox(0,0){$|0\rangle$}}%
% STR 2 0 3 0
% 3 3780 910 3780 1010 5 0
% $\alpha |0\rangle+\beta |1\rangle$
\put(37.8000,-10.1000){\makebox(0,0){$\alpha |0\rangle+\beta |1\rangle$}}%
% LINE 2 0 3 0
% 2 4800 410 5590 1410
% 
\special{pn 8}%
\special{pa 4800 410}%
\special{pa 5590 1410}%
\special{fp}%
% LINE 2 0 3 0
% 2 4800 1380 5590 2380
% 
\special{pn 8}%
\special{pa 4800 1380}%
\special{pa 5590 2380}%
\special{fp}%
% LINE 2 0 3 0
% 2 4800 400 4800 1380
% 
\special{pn 8}%
\special{pa 4800 400}%
\special{pa 4800 1380}%
\special{fp}%
% LINE 2 0 3 0
% 2 5590 1400 5590 2380
% 
\special{pn 8}%
\special{pa 5590 1400}%
\special{pa 5590 2380}%
\special{fp}%
% VECTOR 2 0 3 0
% 2 1200 4400 6810 4400
% 
\special{pn 8}%
\special{pa 1200 4400}%
\special{pa 6810 4400}%
\special{fp}%
\special{sh 1}%
\special{pa 6810 4400}%
\special{pa 6744 4380}%
\special{pa 6758 4400}%
\special{pa 6744 4420}%
\special{pa 6810 4400}%
\special{fp}%
% STR 2 0 3 0
% 3 2200 4530 2200 4630 5 0
% $t=0$
\put(22.0000,-46.3000){\makebox(0,0){$t=0$}}%
% STR 2 0 3 0
% 3 5270 4530 5270 4630 5 0
% $t=t_{0}$
\put(52.7000,-46.3000){\makebox(0,0){$t=t_{0}$}}%
% STR 2 0 3 0
% 3 4550 3180 4550 3280 5 0
% $\alpha |0\rangle +\beta |1\rangle$
\put(45.5000,-32.8000){\makebox(0,0){$\alpha |0\rangle +\beta |1\rangle$}}%
% STR 2 0 3 0
% 3 6360 2710 6360 2810 5 0
% $|0\rangle$ ($\mbox{probability}\ |\alpha|^{2}$)
\put(63.6000,-28.1000){\makebox(0,0){$|0\rangle$ ($\mbox{probability}\ |\alpha|^{2}$)}}%
% STR 2 0 3 0
% 3 6380 3670 6380 3770 5 0
% $|1\rangle$ ($\mbox{probability}\ |\beta|^{2}$)
\put(63.8000,-37.7000){\makebox(0,0){$|1\rangle$ ($\mbox{probability}\ |\beta|^{2}$)}}%
% LINE 2 1 3 0
% 2 2200 1400 5210 1400
% 
\special{pn 8}%
\special{pa 2200 1400}%
\special{pa 5210 1400}%
\special{da 0.070}%
% VECTOR 2 0 3 0
% 2 5010 3260 5480 2910
% 
\special{pn 8}%
\special{pa 5010 3260}%
\special{pa 5480 2910}%
\special{fp}%
\special{sh 1}%
\special{pa 5480 2910}%
\special{pa 5416 2934}%
\special{pa 5438 2942}%
\special{pa 5438 2966}%
\special{pa 5480 2910}%
\special{fp}%
% VECTOR 2 0 3 0
% 2 5010 3310 5480 3660
% 
\special{pn 8}%
\special{pa 5010 3310}%
\special{pa 5480 3660}%
\special{fp}%
\special{sh 1}%
\special{pa 5480 3660}%
\special{pa 5438 3604}%
\special{pa 5438 3628}%
\special{pa 5416 3636}%
\special{pa 5480 3660}%
\special{fp}%
% STR 2 0 3 0
% 3 5960 1380 5960 1480 2 0
% Detector
\put(59.6000,-14.8000){\makebox(0,0)[lb]{Detector}}%
% STR 2 0 3 0
% 3 6940 4350 6940 4450 2 0
% time
\put(69.4000,-44.5000){\makebox(0,0)[lb]{time}}%
% STR 2 0 3 0
% 3 4180 5200 4180 5300 5 0
% Figure I : Image of the Copenhagen interpretation
\put(41.8000,-53.0000){\makebox(0,0){Figure I : Image of the Copenhagen interpretation}}%
% LINE 2 2 3 0
% 2 2200 1390 2200 4400
% 
\special{pn 8}%
\special{pa 2200 1390}%
\special{pa 2200 4400}%
\special{dt 0.045}%
% LINE 2 2 3 0
% 2 5210 1400 5210 4400
% 
\special{pn 8}%
\special{pa 5210 1400}%
\special{pa 5210 4400}%
\special{dt 0.045}%
\end{picture}%

\end{center}
\vspace{5mm}

Here is an important comment. Beginners of QM  might think 
that a quantum state created by an experiment would undergo 
the unitary time evolution (U) forever. 

This is nothing but an illusion because the quantum state is in 
an environment (a kind of heat bath) 
and the interaction with it will disturb the quantum state. 
For example, readers should imagine an oscillator on the desk. 

In order to understand QM deeply readers should take 
decoherence (: interaction with environment) into consideration 
correctly. For this topic see for example \cite{WZ}.

In this paper we try to prove the Copenhagen interpretation 
from the view point of decoherence \footnote{
As far as I know this is a very promising method}. 
Namely, we consider that measurement is a kind of 
{\bf decoherence forced}. 

For the purpose we introduce a decoherence time $t_{D}$, 
which is not necessarily definite. 
The quantum coherence of our system will collapse completely 
when $t>t_{D}$. Therefore, we must finish measuring 
the system within $t_{D}$ ($t_{0}\ll t_{D}$).

\vspace{5mm}
\begin{center}
%WinTpicVersion3.08
\unitlength 0.1in
\begin{picture}( 35.7500,  9.1000)( 20.4500,-18.1500)
% LINE 2 0 3 0
% 2 2190 1080 2190 1320
% 
\special{pn 8}%
\special{pa 2190 1080}%
\special{pa 2190 1320}%
\special{fp}%
% VECTOR 2 0 3 0
% 4 2210 1260 2210 1260 5620 1260 5620 1260
% 
\special{pn 8}%
\special{pa 2210 1260}%
\special{pa 2210 1260}%
\special{fp}%
\special{pa 5620 1260}%
\special{pa 5620 1260}%
\special{fp}%
% VECTOR 2 0 3 0
% 4 2210 1195 2210 1195 2200 1195 5220 1195
% 
\special{pn 8}%
\special{pa 2210 1196}%
\special{pa 2210 1196}%
\special{fp}%
\special{pa 2200 1196}%
\special{pa 5220 1196}%
\special{fp}%
\special{sh 1}%
\special{pa 5220 1196}%
\special{pa 5154 1176}%
\special{pa 5168 1196}%
\special{pa 5154 1216}%
\special{pa 5220 1196}%
\special{fp}%
% STR 2 0 3 0
% 3 3210 1330 3210 1430 5 0
% $t_{D}$
\put(32.1000,-14.3000){\makebox(0,0){$t_{D}$}}%
% STR 2 0 3 0
% 3 2500 1330 2500 1430 5 0
% $t_{0}$
\put(25.0000,-14.3000){\makebox(0,0){$t_{0}$}}%
% STR 2 0 3 0
% 3 5520 1080 5520 1180 5 0
% {time}
\put(55.2000,-11.8000){\makebox(0,0){{time}}}%
% STR 2 0 3 0
% 3 2180 1320 2180 1420 5 0
% $0$
\put(21.8000,-14.2000){\makebox(0,0){$0$}}%
% STR 2 0 3 0
% 3 4150 890 4150 990 5 0
% decoherence
\put(41.5000,-9.9000){\makebox(0,0){decoherence}}%
% DOT 2 0 3 0
% 2 2500 1190 2500 1190
% 
\special{pn 8}%
\special{sh 1}%
\special{ar 2500 1190 10 10 0  6.28318530717959E+0000}%
\special{sh 1}%
\special{ar 2500 1190 10 10 0  6.28318530717959E+0000}%
% DOT 2 0 3 0
% 2 3200 1190 3200 1190
% 
\special{pn 8}%
\special{sh 1}%
\special{ar 3200 1190 10 10 0  6.28318530717959E+0000}%
\special{sh 1}%
\special{ar 3200 1190 10 10 0  6.28318530717959E+0000}%
% STR 2 0 3 0
% 3 3750 1800 3750 1900 5 0
% Figure II : Decoherence time
\put(37.5000,-19.0000){\makebox(0,0){Figure II : Decoherence time}}%
\end{picture}%

\end{center}
\vspace{5mm}

\section{``Proof" of the Copenhagen Interpretation}
In this section we try to give a proof to the Copenhagen 
Interpretation. We perform this by embedding it into 
decoherence theory. The method developped in the following 
is based on the paper \cite{KF3}.

\subsection{General Theory}
We consider an atom flying as in the figure of the preceding 
section and treat a two level system of the atom in the following, 
see for example \cite{WS}. 
First of all let us prepare some notations from Quantum Optics. 
Since we treat the two level system of the atom 
the target space is  
${\bf C}^{2}=\mbox{Vect}_{{\bf C}}(|{0}\rangle, |{1}\rangle)$ 
with bases
\[
|{0}\rangle=
\left(
\begin{array}{c}
1 \\
0
\end{array}
\right),\quad
|{1}\rangle=
\left(
\begin{array}{c}
0 \\
1
\end{array}
\right).
\]
Then Pauli matrices $\{\sigma_{1},\ \sigma_{2},\ \sigma_{3}\}$ 
with the identity $1_{2}$
\[
\sigma_{1}=
\left(
\begin{array}{cc}
0 & 1 \\
1 & 0
\end{array}
\right),\quad
\sigma_{2}=
\left(
\begin{array}{cc}
0 & -i \\
i  & 0
\end{array}
\right),\quad
\sigma_{3}=
\left(
\begin{array}{cc}
1 & 0  \\
0 & -1
\end{array}
\right),\quad
1_{2}=
\left(
\begin{array}{cc}
1 & 0 \\
0 & 1
\end{array}
\right)
\]
act on the space. For
\[
\sigma_{+}\equiv \frac{1}{2}(\sigma_{1}+i\sigma_{2})=
\left(
\begin{array}{cc}
0 & 1 \\
0 & 0
\end{array}
\right),\quad
\sigma_{-}\equiv \frac{1}{2}(\sigma_{1}-i\sigma_{2})=
\left(
\begin{array}{cc}
0 & 0 \\
1 & 0
\end{array}
\right)
\]
it is easy to see
\[
\sigma_{+}\sigma_{-}=
\left(
\begin{array}{cc}
1 & 0 \\
0 & 0
\end{array}
\right),\quad
\sigma_{-}\sigma_{+}=
\left(
\begin{array}{cc}
0 & 0 \\
0 & 1
\end{array}
\right).
\]

Here we may assume that the initial state is $\ket{0}$ at $t=0$ 
and the intermediate state is $\alpha\ket{0}+\beta\ket{1}$ 
for $0<t<t_{0}$ and the last state is the one detected at 
$t=t_{0}$, see the figure in the preceding section once more.

For the initial time $t=0$ we can assume that the Hamiltonian 
is a diagonal form
\begin{equation}
\label{eq:diagonal matrix}
H_{0}=
\left(
\begin{array}{cc}
E_{0} & 0  \\
0 & E_{1}
\end{array}
\right)
\end{equation}
where $E_{0}$ and $E_{1}$ are the eigenvalues 
($E_{0}<E_{1}$ for simplicity) of the atom. It is easy to see
\[
H_{0}\ket{0}=E_{0}\ket{0},\quad H_{0}\ket{1}=E_{1}\ket{1}.
\]

By subjecting a laser field to the atom (at $t=0_{+}$) 
we can take the Hamiltonian to be 
\begin{eqnarray}
\label{eq:entangled form}
H
&=&
U(\alpha,\beta)
\left(
\begin{array}{cc}
E_{0} & 0  \\
0 & E_{1}
\end{array}
\right)
{U(\alpha,\beta)}^{\dagger} \nonumber \\
&=&
\left(
\begin{array}{cc}
|\alpha|^{2}E_{0}+|\beta|^{2}E_{1} & \alpha\bar{\beta}(E_{0}-E_{1}) \\
\bar{\alpha}\beta(E_{0}-E_{1}) & |\beta|^{2}E_{0}+|\alpha|^{2}E_{1}
\end{array}
\right)
\end{eqnarray}
for the intermediate time $0<t<t_{0}$, where $U(\alpha,\beta)$ is 
a special unitary matrix given by 
\begin{equation}
\label{eq:unitary matrix}
U=U(\alpha,\beta)=
\left(
\begin{array}{cc}
\alpha & -\bar{\beta}  \\
\beta & \bar{\alpha}
\end{array}
\right)\quad (|\alpha|^{2}+|\beta|^{2}=1).
\end{equation}
In this case, it is easy to see
\[
\alpha|{0}\rangle +\beta|{1}\rangle
=
\left(
\begin{array}{c}
\alpha \\
\beta
\end{array}
\right), 
\quad 
-\bar{\beta}|{0}\rangle +\bar{\alpha}|{1}\rangle
=
\left(
\begin{array}{c}
-\bar{\beta} \\
\bar{\alpha}
\end{array}
\right)
\]
and
\[
H(\alpha|{0}\rangle +\beta|{1}\rangle)=E_{0}(\alpha|{0}\rangle +\beta|{1}\rangle), \quad
H(-\bar{\beta}|{0}\rangle +\bar{\alpha}|{1}\rangle)=E_{1}(-\bar{\beta}|{0}\rangle +\bar{\alpha}|{1}\rangle).
\]
Note that $H$ and $H_{0}$ are of course hermitian matrices 
($H=H^{\dagger}$,\ $H_{0}=H_{0}^{\dagger}$).

\vspace{3mm}
Since
\[
(\alpha|{0}\rangle +\beta|{1}\rangle)(\alpha|{0}\rangle +\beta|{1}\rangle)^{\dagger}
=
|\alpha|^{2}\ket{0}\bra{0}+\alpha\bar{\beta}\ket{0}\bra{1}+
\bar{\alpha}\beta\ket{1}\bra{0}+|\beta|^{2}\ket{1}\bra{1}
\]
the Copenhagen interpretation may be written as collapsing
\[
(\alpha|{0}\rangle +\beta|{1}\rangle)(\alpha|{0}\rangle +\beta|{1}\rangle)^{\dagger}
\longrightarrow 
|\alpha|^{2}\ket{0}\bra{0}+|\beta|^{2}\ket{1}\bra{1}.
\]

To treat decoherence in a correct manner we must change 
models based on from a pure state to a density matrix. 
The general definition of density matrix $\rho$ is given by both 
$\rho^{\dagger}=\rho$ and $\tr{\rho}=1$, so we can write 
$\rho=\rho(t)$ as
\begin{equation}
\label{eq:density matrix}
\rho=
\left(
\begin{array}{cc}
a         & b  \\
\bar{b} & d
\end{array}
\right)
\quad (a=\bar{a},\ d=\bar{d},\ a+d=1).
\end{equation}

The general form of master equation (\cite{Lind}, \cite{GKS})\footnote{
In standard textbooks of QM  decoherence theory is usually not 
contained, so it may be difficult to beginners (young students). 
See for example \cite{BP} or \cite{KH}} 
is well--known to be
\begin{equation}
\label{eq:master equation}
\frac{d}{dt}\rho=-i[H, \rho]+D\rho \quad (\Leftarrow \hbar=1\ 
\mbox{for simplicity})
\end{equation}
where
\[
D\rho=
\mu\left(\sigma_{-}\rho\sigma_{+}-\frac{1}{2}\sigma_{+}\sigma_{-}\rho-\frac{1}{2}\rho\sigma_{+}\sigma_{-}\right)
+
\nu\left(\sigma_{+}\rho\sigma_{-}-\frac{1}{2}\sigma_{-}\sigma_{+}\rho-\frac{1}{2}\rho\sigma_{-}\sigma_{+}\right)
\]
and $\mu,\ \nu>0$. Note that $\mu$ and $\nu$ are important 
constants determined later.

We must solve the equation. 
If we write $H$ in (\ref{eq:entangled form}) as
\begin{equation}
\label{eq:entangled form 2}
H=
\left(
\begin{array}{cc}
h        & k \\
\bar{k} & l
\end{array}
\right)
\quad (h,\ l \in \real, \ k \in \fukuso)
\end{equation}
for simplicity, then the master equation above can be rewritten as
\begin{equation}
\label{eq:master equation 2}
\frac{d}{dt}
\left(
\begin{array}{c}
a \\
b \\
\bar{b} \\
d
\end{array}
\right)
=
\left(
\begin{array}{cccc}
-\mu & i\bar{k} & -ik & \nu                                \\
ik & i(l-h)-\frac{\mu+\nu}{2} & 0 & -ik                  \\
-i\bar{k} & 0 & -i(l-h)-\frac{\mu+\nu}{2} & i\bar{k} \\
\mu & -i\bar{k} & ik & -\nu 
\end{array}
\right)
\left(
\begin{array}{c}
a \\
b \\
\bar{b} \\
d
\end{array}
\right).
\end{equation}
The derivation is left to readers.

Note and set
\begin{eqnarray*}
&&\left(
\begin{array}{cccc}
-\mu & i\bar{k} & -ik & \nu                               \\
ik & i(l-h)-\frac{\mu+\nu}{2} & 0 & -ik                  \\
-i\bar{k} & 0 & -i(l-h)-\frac{\mu+\nu}{2} & i\bar{k} \\
\mu & -i\bar{k} & ik & -\nu 
\end{array}
\right) \\
&=&
\left(
\begin{array}{cccc}
0 & i\bar{k} & -ik & 0               \\
ik & i(l-h) & 0 & -ik                  \\
-i\bar{k} & 0 & -i(l-h) & i\bar{k}  \\
0 & -i\bar{k} & ik & 0
\end{array}
\right)
+
\left(
\begin{array}{cccc}
-\mu & 0 & 0 & \nu                 \\
0 & -\frac{\mu+\nu}{2} & 0 & 0  \\
0 & 0 & -\frac{\mu+\nu}{2} & 0  \\
\mu & 0 & 0 & -\nu 
\end{array}
\right) \\
&\equiv& \widehat{H}+\widehat{D}.
\end{eqnarray*}
The general solution of (\ref{eq:master equation 2}) 
is given by
\begin{equation}
\label{eq:general solution}
\left(
\begin{array}{c}
a(t) \\
b(t) \\
\bar{b}(t) \\
d(t)
\end{array}
\right)
=
e^{t\left(\widehat{H}+\widehat{D}\right)}
\left(
\begin{array}{c}
a(0) \\
b(0) \\
\bar{b}(0) \\
d(0)
\end{array}
\right).
\end{equation}

However, it is not easy to calculate the term 
$e^{t\left(\widehat{H}+\widehat{D}\right)}$ exactly, so 
we use some approximation. 
In general, the Zassenhaus formula (see for example 
\cite{CZ}, \cite{Five}) is convenient

\vspace{3mm}\noindent
{\bf Zassenhaus Formula}\ \ For operators (or square matrices) 
$A$ and $B$ we have an expansion
\begin{equation}
\label{eq:Zassenhaus formula}
e^{t(A+B)}=
\cdots 
e^{-\frac{t^{3}}{6}\{2[[A,B],B]+[[A,B],A]\}}
e^{\frac{t^{2}}{2}[A,B]}
e^{tB}
e^{tA}.
\end{equation}
The proof is easy. Up to $O(t^{2})$ we obtain

\begin{eqnarray*}
e^{\frac{t^{2}}{2}[A,B]}e^{tB}e^{tA}
&=&
\left({\bf 1}+\frac{t^{2}}{2}[A,B]\right)
\left({\bf 1}+tB+\frac{t^{2}}{2}B^{2}\right)
\left({\bf 1}+tA+\frac{t^{2}}{2}A^{2}\right) \\
&=&
\left({\bf 1}+\frac{t^{2}}{2}(AB-BA)\right)
\left({\bf 1}+t(A+B)+\frac{t^{2}}{2}(A^{2}+2BA+B^{2})\right) \\
&=&
{\bf 1}+t(A+B)+\frac{t^{2}}{2}(A^{2}+2BA+B^{2}+AB-BA) \\
&=&
{\bf 1}+t(A+B)+\frac{t^{2}}{2}(A^{2}+AB+BA+B^{2}) \\
&=&
{\bf 1}+t(A+B)+\frac{t^{2}}{2}(A+B)^{2} \\
&=&
e^{t(A+B)}.
\end{eqnarray*}
To check the equation up to $O(t^{3})$ is left to readers, 
which is a good exercise for undergraduates.

Note that the formula is a bit different from that of \cite{CZ}. 
Zassenhaus formula is a kind of converse of the 
Baker-Campbell-Hausdorff formula
\[
e^{A}e^{B}=e^{A+B+\frac{1}{2}[A,B]+\frac{1}{12}\{[[A,B],B]+
[[B,A],A]\}+\cdots}
\]
where $t=1$ for simplicity.

\subsection{Measurement (= Decoherence Forced)}
The decoherence time $t_{D}$ is in general very short and 
the measurement must be performed within the time 
($0<t_{0}<t_{D}$). From this 
the essential part of $e^{t\left(\widehat{H}+\widehat{D}\right)}$ is 
\[
e^{t\left(\widehat{H}+\widehat{D}\right)}\approx 
e^{t\widehat{D}}e^{t\widehat{H}}
\]
for $0<t<t_{0}$. 

To embed the measurement (: decoherence forced) 
into decoherence theory means that 
we treat the approximate solution
\begin{equation}
\label{eq:general approximate solution}
\left(
\begin{array}{c}
a(t) \\
b(t) \\
\bar{b}(t) \\
d(t)
\end{array}
\right)
\approx 
e^{t\widehat{D}}e^{t\widehat{H}}
\left(
\begin{array}{c}
a(0) \\
b(0) \\
\bar{b}(0) \\
d(0)
\end{array}
\right)
\quad (t\geq 0)
\end{equation}
instead of treating the full solution (\ref{eq:general solution}). 
See the following figure.

\vspace{5mm}
\begin{center}
%WinTpicVersion4.30a
{\unitlength 0.1in%
\begin{picture}( 41.2000, 16.2000)( 13.4000,-21.3500)%
% STR 2 0 3 0 Black White
% 4 3620 1670 3620 1770 5 0 0 0
% $t_{D}$
\put(36.2000,-17.7000){\makebox(0,0){$t_{D}$}}%
% LINE 2 0 3 0 Black White
% 2 2000 900 2000 1110
% 
\special{pn 8}%
\special{pa 2000 900}%
\special{pa 2000 1110}%
\special{fp}%
% LINE 2 0 3 0 Black White
% 2 2000 1500 2000 1710
% 
\special{pn 8}%
\special{pa 2000 1500}%
\special{pa 2000 1710}%
\special{fp}%
% VECTOR 2 0 3 0 Black White
% 2 2000 1000 5220 1000
% 
\special{pn 8}%
\special{pa 2000 1000}%
\special{pa 5220 1000}%
\special{fp}%
\special{sh 1}%
\special{pa 5220 1000}%
\special{pa 5153 980}%
\special{pa 5167 1000}%
\special{pa 5153 1020}%
\special{pa 5220 1000}%
\special{fp}%
% VECTOR 2 0 3 0 Black White
% 2 2000 1600 5220 1600
% 
\special{pn 8}%
\special{pa 2000 1600}%
\special{pa 5220 1600}%
\special{fp}%
\special{sh 1}%
\special{pa 5220 1600}%
\special{pa 5153 1580}%
\special{pa 5167 1600}%
\special{pa 5153 1620}%
\special{pa 5220 1600}%
\special{fp}%
% STR 2 0 3 0 Black White
% 4 2610 740 2610 840 5 0 0 0
% $t_{0}$
\put(26.1000,-8.4000){\makebox(0,0){$t_{0}$}}%
% DOT 2 0 3 0 Black White
% 2 2600 1000 2600 1000
% 
\special{pn 4}%
\special{sh 1}%
\special{ar 2600 1000 8 8 0  6.28318530717959E+0000}%
\special{sh 1}%
\special{ar 2600 1000 8 8 0  6.28318530717959E+0000}%
% STR 2 0 3 0 Black White
% 4 3600 750 3600 850 5 0 0 0
% $t_{D}$
\put(36.0000,-8.5000){\makebox(0,0){$t_{D}$}}%
% DOT 2 0 3 0 Black White
% 2 3600 1000 3600 1000
% 
\special{pn 4}%
\special{sh 1}%
\special{ar 3600 1000 8 8 0  6.28318530717959E+0000}%
\special{sh 1}%
\special{ar 3600 1000 8 8 0  6.28318530717959E+0000}%
% DOT 2 0 3 0 Black White
% 2 3600 1600 3600 1600
% 
\special{pn 4}%
\special{sh 1}%
\special{ar 3600 1600 8 8 0  6.28318530717959E+0000}%
\special{sh 1}%
\special{ar 3600 1600 8 8 0  6.28318530717959E+0000}%
% STR 2 0 3 0 Black White
% 4 2290 480 2290 580 5 0 0 0
% $e^{t\widehat{D}}e^{t\widehat{H}}$
\put(22.9000,-5.8000){\makebox(0,0){$e^{t\widehat{D}}e^{t\widehat{H}}$}}%
% STR 2 0 3 0 Black White
% 4 4350 480 4350 580 5 0 0 0
% $e^{t(\widehat{H}+\widehat{D})}$
\put(43.5000,-5.8000){\makebox(0,0){$e^{t(\widehat{H}+\widehat{D})}$}}%
% VECTOR 2 0 3 0 Black White
% 2 2320 1100 4380 1490
% 
\special{pn 8}%
\special{pa 2320 1100}%
\special{pa 4380 1490}%
\special{fp}%
\special{sh 1}%
\special{pa 4380 1490}%
\special{pa 4318 1458}%
\special{pa 4328 1480}%
\special{pa 4311 1497}%
\special{pa 4380 1490}%
\special{fp}%
% STR 2 0 3 0 Black White
% 4 3640 2100 3640 2200 5 0 0 0
% Figure III : Embedding of the measurement into decoherence theory
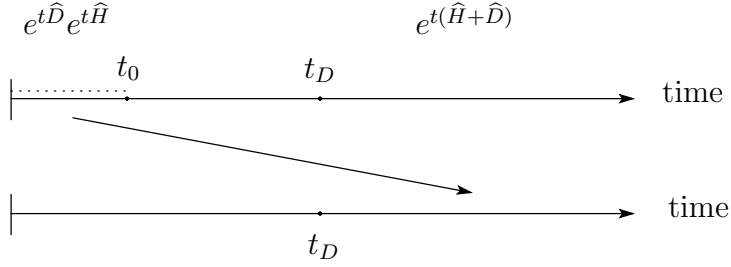
\put(36.4000,-22.0000){\makebox(0,0){Figure III : Embedding of the measurement into decoherence theory}}%
% STR 2 0 3 0 Black White
% 4 5530 860 5530 960 5 0 0 0
% time
\put(55.3000,-9.6000){\makebox(0,0){time}}%
% STR 2 0 3 0 Black White
% 4 5560 1460 5560 1560 5 0 0 0
% time
\put(55.6000,-15.6000){\makebox(0,0){time}}%
% LINE 2 2 3 0 Black White
% 2 2000 960 2600 960
% 
\special{pn 8}%
\special{pa 2000 960}%
\special{pa 2600 960}%
\special{dt 0.045}%
\end{picture}}%

\end{center}
\vspace{5mm}

First, let us calculate $e^{t\widehat{D}}$. For the purpose 
we set
\[
K=
\left(
\begin{array}{cc}
-\mu & \nu \\
\mu & -\nu 
\end{array}
\right)
\]
and calculate $e^{tK}$. The eigenvalues of $K$ are 
$\{0,-(\mu+\nu)\}$ and corresponding eigenvectors (
not normalized) are
\[
0\longleftrightarrow 
\left(
\begin{array}{c}
\nu \\
\mu 
\end{array}
\right),\quad
-(\mu+\nu)\longleftrightarrow 
\left(
\begin{array}{c}
1  \\
-1
\end{array}
\right).
\]
If we define the matrix
\[
O=
\left(
\begin{array}{cc}
\nu & 1   \\
\mu & -1
\end{array}
\right)
\Longrightarrow 
O^{-1}=\frac{1}{\mu+\nu}
\left(
\begin{array}{cc}
1 & 1          \\
\mu & -\nu 
\end{array}
\right)
\]
then it is easy to see
\[
K=
O
\left(
\begin{array}{cc}
0 &                 \\
  & -(\mu+\nu)
\end{array}
\right)
O^{-1}
\]
and
\[
e^{tK}=
O
\left(
\begin{array}{cc}
1 &                         \\
   & e^{-t(\mu+\nu)}
\end{array}
\right)
O^{-1}
=
\frac{1}{\mu+\nu}
\left(
\begin{array}{cc}
\nu+\mu e^{-t(\mu+\nu)} & \nu-\nu e^{-t(\mu+\nu)}  \\
\mu-\mu e^{-t(\mu+\nu)} & \mu+\nu e^{-t(\mu+\nu)}
\end{array}
\right).
\]
Therefore, we have
\begin{equation}
\label{eq:E(tD)}
e^{t\widehat{D}}
=
\left(
\begin{array}{cccc}
\frac{\nu+\mu e^{-t(\mu+\nu)}}{\mu+\nu} & 0 & 0 & \frac{\nu-\nu e^{-t(\mu+\nu)}}{\mu+\nu}  \\
0 & e^{-t\frac{\mu+\nu}{2}} & 0 & 0                                                                                  \\
0 & 0 & e^{-t\frac{\mu+\nu}{2}} & 0                                                                                  \\
\frac{\mu-\mu e^{-t(\mu+\nu)}}{\mu+\nu} & 0 & 0 & \frac{\mu+\nu e^{-t(\mu+\nu)}}{\mu+\nu}
\end{array}
\right)
\approx
\frac{1}{\mu+\nu}
\left(
\begin{array}{cccc}
\nu & 0 & 0 & \nu   \\
0 & 0 & 0  & 0        \\
0 & 0 & 0  & 0        \\
\mu & 0 & 0 & \mu 
\end{array}
\right) 
\end{equation}
if $t$ is large enough ($t\gg 1/(\mu+\nu)$).

Next, let us calculate $e^{t\widehat{H}}$. Since we need 
some properties of tensor product in the following see 
for example \cite{Five}. We can write the equation as
\[
\widehat{H}=-i\left(H\otimes 1_{2}-1_{2}\otimes H^{T}\right).
\]
In fact,
\begin{eqnarray*}
\widehat{H}
&=&-i
\left\{
\left(
\begin{array}{cc}
h        & k \\
\bar{k} & l
\end{array}
\right)
\otimes
\left(
\begin{array}{cc}
1 & 0 \\
0 & 1
\end{array}
\right)
-
\left(
\begin{array}{cc}
1 & 0 \\
0 & 1
\end{array}
\right)
\otimes
\left(
\begin{array}{cc}
h & \bar{k} \\
k & l
\end{array}
\right)
\right\} \\
&=&-i
\left\{
\left(
\begin{array}{cccc}
h & 0 & k & 0 \\
0 & h & 0 & k \\
\bar{k} & 0 & l  & 0 \\
0 & \bar{k} & 0 & l
\end{array}
\right)
-
\left(
\begin{array}{cccc}
h & \bar{k} & 0 & 0 \\
k & l  & 0 & 0 \\
0 & 0 & h & \bar{k} \\
0 & 0 & k & l 
\end{array}
\right)
\right\} \\
&=&-i
\left(
\begin{array}{cccc}
0 & -\bar{k} & k & 0  \\
-k & -(l-h) & 0 & k \\
\bar{k} & 0 & l-h & -\bar{k} \\
0 & \bar{k} & -k & 0
\end{array}
\right).
\end{eqnarray*}

It is well--known that
\[
e^{t\widehat{H}}
=
e^{-it\left(H\otimes 1_{2}-1_{2}\otimes H^{T}\right)}
=
e^{-it H\otimes 1_{2}}e^{it 1_{2}\otimes H^{T}}
=
\left(e^{-itH}\otimes 1_{2}\right)
\left(1_{2}\otimes e^{itH^{T}}\right)
=
e^{-itH}\otimes e^{itH^{T}}.
\]
Since
\[
H=
U
\left(
\begin{array}{cc}
E_{0} & 0 \\
0 & E_{1}
\end{array}
\right)
U^{\dagger}
\]
we have
\[
e^{-itH}=
U
\left(
\begin{array}{cc}
e^{-itE_{0}} & 0 \\
0 & e^{-itE_{1}}
\end{array}
\right)
U^{\dagger}
\]
and
\begin{eqnarray*}
e^{t\widehat{H}}
&=&
\left\{
U
\left(
\begin{array}{cc}
e^{-itE_{0}} & 0 \\
0 & e^{-itE_{1}}
\end{array}
\right)
U^{\dagger}
\right\}
\otimes
\left\{
U
\left(
\begin{array}{cc}
e^{itE_{0}} & 0 \\
0 & e^{itE_{1}}
\end{array}
\right)
U^{\dagger}
\right\}^{T} \\
&=&
\left\{
U
\left(
\begin{array}{cc}
e^{-itE_{0}} & 0 \\
0 & e^{-itE_{1}}
\end{array}
\right)
U^{\dagger}
\right\}
\otimes
\left\{
(U^{\dagger})^{T}
\left(
\begin{array}{cc}
e^{itE_{0}} & 0 \\
0 & e^{itE_{1}}
\end{array}
\right)
U^{T}
\right\} \\
&=&
U\otimes (U^{\dagger})^{T}
\left\{
\left(
\begin{array}{cc}
e^{-itE_{0}} & 0 \\
0 & e^{-itE_{1}}
\end{array}
\right)
\otimes
\left(
\begin{array}{cc}
e^{itE_{0}} & 0 \\
0 & e^{itE_{1}}
\end{array}
\right)
\right\}
\left(U\otimes (U^{\dagger})^{T}\right)^{\dagger} \\
&=&
U\otimes (U^{\dagger})^{T}
\left(
\begin{array}{cccc}
1 &  &  &                         \\
  & e^{it(E_{1}-E_{0})} &  &    \\
  &   & e^{-it(E_{1}-E_{0})} & \\
  &   &   & 1
\end{array}
\right)
\left(U\otimes (U^{\dagger})^{T}\right)^{\dagger}.
\end{eqnarray*}
Here we have used well--known formulas on tensor 
product
\begin{eqnarray*}
&&
(A_{1}\otimes B_{1})(A_{2}\otimes B_{2})=A_{1}A_{2}\otimes B_{1}B_{2},\ \ 
(A_{1}\otimes B_{1})(A_{2}\otimes B_{2})(A_{3}\otimes B_{3})
=A_{1}A_{2}A_{3}\otimes B_{1}B_{2}B_{3}, \\
&&
(A\otimes B)^{\dagger}=A^{\dagger}\otimes B^{\dagger},\ 
(A\otimes B)^{T}=A^{T}\otimes B^{T},
\end{eqnarray*}
see for example \cite{Five}.
 
Since
\[
U=U(\alpha,\beta)=
\left(
\begin{array}{cc}
\alpha & -\bar{\beta}  \\
\beta & \bar{\alpha}
\end{array}
\right)\quad (|\alpha|^{2}+|\beta|^{2}=1)
\]
from (\ref{eq:unitary matrix}) we have
\[
U\otimes (U^{\dagger})^{T}
=
\left(
\begin{array}{cccc}
|\alpha|^{2} & -\alpha\beta & -\bar{\alpha}\bar{\beta} & |\beta|^{2}      \\
\alpha\bar{\beta} & \alpha^{2} & -\bar{\beta}^{2} & -\alpha\bar{\beta}  \\
\bar{\alpha}\beta & -\beta^{2} & \bar{\alpha}^{2} & -\bar{\alpha}\beta  \\
|\beta|^{2} & \alpha\beta & \bar{\alpha}\bar{\beta} & |\alpha|^{2}
\end{array}
\right)
\]
and, by setting $J=e^{it(E_{1}-E_{0})}$ for simplicity, 
\begin{eqnarray}
\label{eq:E(tH)}
e^{t\widehat{H}}
&=&
U\otimes (U^{\dagger})^{T}
\left(
\begin{array}{cccc}
1 &  &  &          \\
  & J &  &        \\
  &   & J^{-1} & \\
  &   &   & 1
\end{array}
\right)
\left(U\otimes (U^{\dagger})^{T}\right)^{\dagger} \nonumber \\
&=&
\left(
\begin{array}{cccc}
c_{11} & c_{12} & c_{13} & c_{14}  \\
* & * & * & *                                                     \\
* & * & * & *                                                     \\
c_{41} & c_{42} & c_{43} & c_{44}
\end{array}
\right)
\end{eqnarray}
where
\begin{eqnarray*}
c_{11}&=&|\alpha|^{4}+(J+J^{-1})|\alpha|^{2}|\beta|^{2}+|\beta|^{4}, \\
c_{12}&=&\left(|\alpha|^{2}-|\alpha|^{2}J+|\beta|^{2}J^{-1}-|\beta|^{2}\right)
\bar{\alpha}\beta, \\
c_{13}&=&\left(|\alpha|^{2}+|\beta|^{2}J-|\alpha|^{2}J^{-1}-|\beta|^{2}\right)
\alpha\bar{\beta}, \\
c_{14}&=&(2-J-J^{-1})|\alpha|^{2}|\beta|^{2}
\end{eqnarray*}
and
\begin{eqnarray*}
c_{41}&=&(2-J-J^{-1})|\alpha|^{2}|\beta|^{2}, \\
c_{42}&=&\left(|\beta|^{2}+|\alpha|^{2}J-|\beta|^{2}J^{-1}-|\alpha|^{2}\right)
\bar{\alpha}\beta, \\
c_{43}&=&\left(|\beta|^{2}-|\beta|^{2}J+|\alpha|^{2}J^{-1}-|\alpha|^{2}\right)
\alpha\bar{\beta}, \\
c_{44}&=&|\beta|^{4}+(J+J^{-1})|\alpha|^{2}|\beta|^{2}+|\alpha|^{4}.
\end{eqnarray*}
Note that $*$'s in the matrix are elements not used in the latter. 
The derivation is left to readers. 

\noindent
Here, we list very important relations among $\{\alpha\}$ 
(coming from $|\alpha|^{2}+|\beta|^{2}=1$)
\begin{equation}
\label{eq: important relations}
c_{11}+c_{41}=1,\quad
c_{12}+c_{42}=0,\quad
c_{13}+c_{43}=0,\quad
c_{14}+c_{44}=1.
\end{equation}

Therefore, from (\ref{eq:general approximate solution}), 
(\ref{eq:E(tD)}), (\ref{eq:E(tH)}) and 
(\ref{eq: important relations}) we obtain
\begin{eqnarray}
\label{eq:Fujii}
\left(
\begin{array}{c}
a(t) \\
b(t) \\
\bar{b}(t) \\
d(t)
\end{array}
\right)
&\approx& 
\frac{1}{\mu+\nu}
\left(
\begin{array}{cccc}
\nu & 0 & 0 & \nu   \\
0 & 0 & 0  & 0        \\
0 & 0 & 0  & 0        \\
\mu & 0 & 0 & \mu 
\end{array}
\right)
\left(
\begin{array}{cccc}
c_{11} & c_{12} & c_{13} & c_{14}  \\
* & * & * & *                         \\
* & * & * & *                         \\
c_{41} & c_{42} & c_{43} & c_{44}
\end{array}
\right)
\left(
\begin{array}{c}
a(0) \\
b(0) \\
\bar{b}(0) \\
d(0)
\end{array}
\right) \nonumber \\
&=& 
\frac{1}{\mu+\nu}
\left(
\begin{array}{cccc}
\nu & 0 & 0 & \nu   \\
0 & 0 & 0  & 0        \\
0 & 0 & 0  & 0        \\
\mu & 0 & 0 & \mu 
\end{array}
\right)
\left(
\begin{array}{c}
a(0) \\
b(0) \\
\bar{b}(0) \\
d(0)
\end{array}
\right) 
\end{eqnarray}
for $t\gg 1/(\mu+\nu)$.

From (\ref{eq:density matrix})
\[
\rho(t)=
\left(
\begin{array}{cc}
a(t)         & b(t)  \\
\bar{b}(t) & d(t)
\end{array}
\right)
\]
we have
\[
\rho(\infty)=
\frac{1}{\mu+\nu}
\left(
\begin{array}{cc}
\nu\left(a(0)+d(0)\right) & 0  \\
0 & \mu\left(a(0)+d(0)\right) 
\end{array}
\right).
\]
The initial density matrix
\[
\rho(0)=\ket{0}\bra{0}
=
\left(
\begin{array}{cc}
1 & 0  \\
0 & 0
\end{array}
\right)
\equiv
\left(
\begin{array}{cc}
a(0)         & b(0)  \\
\bar{b}(0) & d(0)
\end{array}
\right)
\]
gives
\begin{equation}
\label{eq:last form}
\rho(\infty)
=
\frac{1}{\mu+\nu}
\left(
\begin{array}{cc}
\nu & 0  \\
0 & \mu
\end{array}
\right)
=
\frac{\nu}{\mu+\nu}\ket{0}\bra{0}+\frac{\mu}{\mu+\nu}\ket{1}\bra{1}.
\end{equation}
Since
\[
\frac{\nu}{\mu+\nu},\ \frac{\mu}{\mu+\nu}>0 
\quad \mbox{and}\quad 
\frac{\nu}{\mu+\nu}+\frac{\mu}{\mu+\nu}=1
\]
the structure of probability comes out in a natural way.

Moreover, if we can choose $\mu$ and $\nu$ as
\begin{equation}
\frac{\nu}{\mu+\nu}=|\alpha|^{2}
\quad\mbox{and}\quad
\frac{\mu}{\mu+\nu}=|\beta|^{2}
\quad (\Longrightarrow |\alpha|^{2}+|\beta|^{2}=1)
\end{equation}
from the starting point then we have the final form
\begin{equation}
\label{eq:special last form}
\rho(\infty)=|\alpha|^{2}\ket{0}\bra{0}+|\beta|^{2}\ket{1}\bra{1}.
\end{equation}

We can interpret this equation as a mathematical expression 
of the Copenhagen interpretation :\ 
``when a state is superposition $\alpha\ket{0}+\beta\ket{1}$ 
and we observe the observable $Q$ the state collapses like 
$
\alpha\ket{0}+\beta\ket{1}\ \rightarrow \ \ket{0}
\ (\mbox{probability}\ |\alpha|^{2})
\ \ \mbox{or}\ \ 
\alpha\ket{0}+\beta\ket{1}\ \rightarrow \ \ket{1}
\ (\mbox{probability}\ |\beta|^{2})."
$
This finishes the ``proof" of the Copenhagen interpretation.

\vspace{3mm}
The remaining problem is 

\vspace{3mm}\noindent
{\bf Problem}\ \ Why are 
$\frac{\nu}{\mu+\nu}=|\alpha|^{2}$ and $\frac{\mu}{\mu+\nu}=|\beta|^{2}$ 
identified when measuring the system ?

\vspace{3mm}\noindent
It may be difficult to prove the problem without introducing 
another theory. 

\vspace{3mm}
A comment is in order. If we use another approximation
\[
e^{t(\widehat{H}+\widehat{D})}\approx e^{t\widehat{D}}e^{t\widehat{H}}
\quad \Longrightarrow \quad
e^{t(\widehat{H}+\widehat{D})}\approx 
e^{\frac{t^{2}}{2}[\widehat{H}, \widehat{D}]}e^{t\widehat{D}}e^{t\widehat{H}}
\]
we don't have a ``diagonal form" like (\ref{eq:last form}) any more. 
As a result, we can say that in the framework of decoherence theory 
the Copenhagen interpretation is nothing but a special approximate 
phenomenon except for the problem stated above.

\subsection{Decoherence}
Here, we don't observe the system at $t_{0}$ and 
solve the equation (\ref{eq:general solution}) 
\[
\left(
\begin{array}{c}
a(t) \\
b(t) \\
\bar{b}(t) \\
d(t)
\end{array}
\right)
=
e^{t\left(\widehat{H}+\widehat{D}\right)}
\left(
\begin{array}{c}
a(0) \\
b(0) \\
\bar{b}(0) \\
d(0)
\end{array}
\right)
\]
exactly and take the limit $t\ \rightarrow\ \infty$.  

The method is almost equal to that of \cite{KF2}. 
However, since to show it is important as composition 
of the paper, we repeat it within our necessity.

First, we must look for eigenvalues of the matrix $W\equiv 
\widehat{H}+\widehat{D}$
\begin{equation}
\label{eq:W matrix}
W=
\left(
\begin{array}{cccc}
-\mu & i\bar{k} & -ik & \nu                                \\
ik & i(l-h)-\frac{\mu+\nu}{2} & 0 & -ik                  \\
-i\bar{k} & 0 & -i(l-h)-\frac{\mu+\nu}{2} & i\bar{k} \\
\mu & -i\bar{k} & ik & -\nu 
\end{array}
\right).
\end{equation}
For the latter convenience we write the transpose of $W$
\[
W^{T}=
\left(
\begin{array}{cccc}
-\mu & ik & -i\bar{k} & \mu                              \\
i\bar{k} & i(l-h)-\frac{\mu+\nu}{2} & 0 & -i\bar{k}  \\
-ik & 0 & -i(l-h)-\frac{\mu+\nu}{2} & ik               \\
\nu & -ik & i\bar{k} & -\nu 
\end{array}
\right).
\]

Since
\begin{eqnarray*}
0&=&|\lambda 1_{4}-W| \\
&=&
\left|
\begin{array}{cccc}
\lambda+\mu & -i\bar{k} & ik & -\nu                               \\
-ik & \lambda-i(l-h)+\frac{\mu+\nu}{2} & 0 & ik                 \\
i\bar{k} & 0 & \lambda+i(l-h)+\frac{\mu+\nu}{2} & -i\bar{k}  \\
-\mu & i\bar{k} & -ik & \lambda+\nu 
\end{array}
\right| \\
&=& \cdots \\
&=&\lambda
\left|
\begin{array}{cccc}
1 & 0 & 0 & 0                                                                \\
-ik & \lambda-i(l-h)+\frac{\mu+\nu}{2} & 0 & 2ik                 \\
i\bar{k} & 0 & \lambda+i(l-h)+\frac{\mu+\nu}{2} & -2i\bar{k}  \\
-\mu & i\bar{k} & -ik & \lambda+\mu+\nu 
\end{array}
\right| \\
&=&\lambda
\left|
\begin{array}{ccc}
\lambda-i(l-h)+\frac{\mu+\nu}{2} & 0 & 2ik           \\
0 & \lambda+i(l-h)+\frac{\mu+\nu}{2} & -2i\bar{k}  \\
i\bar{k} & -ik & \lambda+\mu+\nu 
\end{array}
\right| \\
&=&\lambda
\left[
\left\{\left(\lambda+\frac{\mu+\nu}{2}\right)^{2}+(l-h)^{2}\right\}(\lambda+\mu+\nu)+
2|k|^{2}(2\lambda+\mu+\nu)
\right]
\end{eqnarray*}
we obtain one trivial root $\lambda=0$ and a cubic equation
\[
\left\{\left(\lambda+\frac{\mu+\nu}{2}\right)^{2}+(l-h)^{2}\right\}(\lambda+\mu+\nu)+
2|k|^{2}(2\lambda+\mu+\nu)=0.
\]
Let us transform this. By setting
\[
\Lambda=\lambda+\frac{\mu+\nu}{2}\ \Longrightarrow\ 
\lambda =\Lambda-\frac{\mu+\nu}{2}
\]
the cubic equation becomes
\[
\Lambda^{3}+\frac{\mu+\nu}{2}\Lambda^{2}+
\{(l-h)^{2}+4|k|^{2}\}\Lambda+(l-h)^{2}\frac{\mu+\nu}{2}=0
\]
and some calculation gives
\begin{equation}
\label{eq:Cubic equation}
\Lambda^{3}+\frac{\mu+\nu}{2}\Lambda^{2}+
(E_{1}-E_{0})^{2}\Lambda+(E_{1}-E_{0})^{2}(|\alpha|^{2}-|\beta|^{2})^{2}\frac{\mu+\nu}{2}=0
\end{equation}
by (\ref{eq:entangled form}) and (\ref{eq:entangled form 2}). 

Here we set
\[
f(\Lambda)=\Lambda^{3}+\frac{\mu+\nu}{2}\Lambda^{2}+
(E_{1}-E_{0})^{2}\Lambda+(E_{1}-E_{0})^{2}(|\alpha|^{2}-|\beta|^{2})^{2}\frac{\mu+\nu}{2}
\]
and treat its roots in an abstract way. 

\vspace{5mm}\noindent
Case (A) : \ $|\alpha|=|\beta|$

In this case
\[
f(\Lambda)=\Lambda\left\{\Lambda^{2}+\frac{\mu+\nu}{2}\Lambda+(E_{1}-E_{0})^{2}\right\},
\]
so we have solutions
\[
\Lambda_{0}=0,\ \ 
\Lambda_{\pm}
=\frac{1}{2}\left\{-\frac{\mu+\nu}{2}\pm \sqrt{\left(\frac{\mu+\nu}{2}\right)^{2}-4(E_{1}-E_{0})^{2}}\right\}.
\]
From these we know
\[
\Lambda_{0}=0,\ \ \Lambda_{\pm}<0
\]
if $\left(\frac{\mu+\nu}{2}\right)^{2}-4(E_{1}-E_{0})^{2}\geq 0$ and
\[
\Lambda_{0}=0,\ \ \mbox{Re}\ \Lambda_{\pm}=-\frac{\mu+\nu}{4} <0
\]
if $\left(\frac{\mu+\nu}{2}\right)^{2}-4(E_{1}-E_{0})^{2}<0$.

\vspace{5mm}\noindent
Case (B) : \ $|\alpha|\ne |\beta|$

We note that $f(\Lambda)>0$ for $\Lambda\geq 0$ because 
all coefficients are positive. Since
\begin{eqnarray*}
f(0)&=&(E_{1}-E_{0})^{2}(|\alpha|^{2}-|\beta|^{2})^{2}\frac{\mu+\nu}{2}>0, \\
f(-\frac{\mu+\nu}{2})&=&-2(E_{1}-E_{0})^{2}(\mu+\nu)|\alpha|^{2}|\beta|^{2}<0
\end{eqnarray*}
there is (at least) one root $-\frac{\mu+\nu}{2}<\Lambda_{0}<0$ 
satisfying $f(\Lambda_{0})=0$. By denoting
\[
f(\Lambda)=\Lambda^{3}+a\Lambda^{2}+b\Lambda+ c
\]
for simplicity we have a decomposition
\[
f(\Lambda)=(\Lambda-\Lambda_{0})
(\Lambda^{2}+(\Lambda_{0}+a)\Lambda +(\Lambda_{0}^{2}+a\Lambda_{0}+b)).
\]
From this we obtain other two roots
\[
\Lambda_{\pm}=\frac{-(\Lambda_{0}+a)\pm 
\sqrt{(\Lambda_{0}+a)^{2}-4(\Lambda_{0}^{2}+a\Lambda_{0}+b)}}{2}.
\]
Note that $\Lambda_{0}+a=\Lambda_{0}+\frac{\mu+\nu}{2}>0$. 
If $\Lambda_{0}^{2}+a\Lambda_{0}+b<0$ then $\Lambda_{+}>0$, 
which is a contradiction. Therefore, $\Lambda_{0}^{2}+a\Lambda_{0}+b>0$.

Therefore,
\[
\Lambda_{0}<0,\ \ \Lambda_{\pm}<0
\]
if $(\Lambda_{0}+a)^{2}-4(\Lambda_{0}^{2}+a\Lambda_{0}+b)>0$ and
\[
\Lambda_{0}<0,\ \ \mbox{Re}\ \Lambda_{\pm}=-\frac{\Lambda_{0}+a}{2}<0.
\]
if $(\Lambda_{0}+a)^{2}-4(\Lambda_{0}^{2}+a\Lambda_{0}+b)<0$. 

As a result, the solutions of the characteristic polynomial of $W$ 
(= $|\lambda 1_{4}-W|$) are
\begin{equation}
\label{eq:solutions}
\lambda_{1}=0,\quad
\lambda_{2}=\Lambda_{0}-\frac{\mu+\nu}{2},\quad
\lambda_{3}=\Lambda_{+}-\frac{\mu+\nu}{2},\quad
\lambda_{4}=\Lambda_{-}-\frac{\mu+\nu}{2}
\end{equation}
and
\begin{equation}
\label{eq:basic properties}
\lambda_{2}<0,\quad \lambda_{3}<0,\quad \lambda_{4}<0
\quad
\mbox{or}
\quad
\lambda_{2}<0,\quad \mbox{Re}\lambda_{3}<0,\quad \mbox{Re}\lambda_{4}<0
\end{equation}
under the conditions stated above.

By the same method in \cite{KF2} : Section 2 we obtain 
the diagonal form
\begin{equation}
\label{eq:diagonal form}
W=(O^{T})^{-1}D_{W}O^{T}
\quad (\Longleftarrow \ W^{T}=OD_{W}O^{-1})
\end{equation}
where $D_{W}$ is the diagonal matrix 
\begin{equation}
\label{eq:diagonal part}
D_{W}=
\left(
\begin{array}{cccc}
0 &                  &                 &                 \\
   & \lambda_{2} &                 &                  \\
   &                 & \lambda_{3} &                  \\
   &                 &                 & \lambda_{4}
\end{array}
\right)
\end{equation}
and $O$ is the matrix consisting of eigenvectors
\begin{equation}
\label{eq:O-1}
O=\left(|0),\ |\lambda_{2}),\ |\lambda_{3}),\ |\lambda_{4})\right)
=
\left(
\begin{array}{cccc}
1 & x_{2} & x_{3} & x_{4}   \\
0 & y_{2} & y_{3} & y_{4}   \\
0 & z_{2} & z_{3} & z_{4}   \\
1 & w_{2} & w_{3} & w_{4}
\end{array}
\right)
\end{equation}
and
\begin{equation}
\label{eq:O-2}
O^{-1}=\frac{1}{|O|}
\left(
\begin{array}{cccc}
\widehat{O}_{11} & \widehat{O}_{12} & \widehat{O}_{13} & \widehat{O}_{14} \\
* & * & * & * \\
* & * & * & * \\
* & * & * & * 
\end{array}
\right)
\end{equation}
where $*$ denotes cofactors unnecessary in the following \footnote{
In order to find the eigenvectors of $W$ it is better to use 
$W^{T}$ rather than $W$ itself. See \cite{KF2} in more detail}.

Here, let us go back to the equation (\ref{eq:master equation 2}).
If we set
\[
(\hat{\rho}=)\Psi=
\left(
\begin{array}{c}
a         \\
b         \\
\bar{b} \\
d
\end{array}
\right)
\]
for simplicity, the equation (\ref{eq:master equation 2}) reads
\[
\frac{d}{dt}\Psi=W\Psi
\]
and the general solution is given by (\ref{eq:diagonal form})
\[
\Psi(t)=e^{tW}\Psi(0)=(O^{T})^{-1}e^{tD_{W}}O^{T}\Psi(0).
\]

Since we are interested in the final state $\Psi(\infty)$ we 
must look for the asymptotic limit 
$\lim_{t\rightarrow \infty}e^{tD_{W}}$.  
It is easy to see
\[
\lim_{t\rightarrow \infty}e^{tD_{W}}
=
\left(
\begin{array}{cccc}
1 & & & \\
& 0 & & \\
& & 0 & \\
& & & 0
\end{array}
\right)
=\kett{0} \braa{0},
\quad
\kett{0}\equiv 
\left(
\begin{array}{c}
1 \\
0 \\
0 \\
0
\end{array}
\right)
\]
by (\ref{eq:basic properties}) and (\ref{eq:diagonal part}), 
so we obtain
\begin{equation}
\label{eq: limit form}
\Psi(\infty)
=(O^{T})^{-1}\kett{0}\braa{0}O^{T}\Psi(0)
=\frac{1}{|O|}
\left(
\begin{array}{cccc}
\widehat{O}_{11} & 0 & 0 & \widehat{O}_{11} \\
\widehat{O}_{12} & 0 & 0 & \widehat{O}_{12} \\
\widehat{O}_{13} & 0 & 0 & \widehat{O}_{13} \\
\widehat{O}_{14} & 0 & 0 & \widehat{O}_{14}
\end{array}
\right)\Psi(0)
\end{equation}
by (\ref{eq:O-1}) and (\ref{eq:O-2}).

This equation gives
\[
\Psi(0)=
\left(
\begin{array}{c}
1 \\
0 \\
0 \\
0
\end{array}
\right)
\ \Longrightarrow \
\Psi(\infty)=\frac{1}{|O|}
\left(
\begin{array}{c}
\widehat{O}_{11} \\
\widehat{O}_{12} \\
\widehat{O}_{13} \\
\widehat{O}_{14}
\end{array}
\right)
\]
and it is equivalent to
\begin{equation}
\label{eq:result 1}
\rho_{0}(0)=\ket{0}\bra{0}=
\left(
\begin{array}{cc}
1 & 0 \\
0 & 0
\end{array}
\right)
\ \Longrightarrow \
\rho_{0}(\infty)=\frac{1}{|O|}
\left(
\begin{array}{cc}
\widehat{O}_{11} & \widehat{O}_{12} \\
\widehat{O}_{13} & \widehat{O}_{14}
\end{array}
\right).
\end{equation}

Similarly,
\[
\Psi(0)=
\left(
\begin{array}{c}
0 \\
0 \\
0 \\
1
\end{array}
\right)
\ \Longrightarrow \
\Psi(\infty)=\frac{1}{|O|}
\left(
\begin{array}{c}
\widehat{O}_{11} \\
\widehat{O}_{12} \\
\widehat{O}_{13} \\
\widehat{O}_{14}
\end{array}
\right)
\]
is equivalent to
\begin{equation}
\label{eq:result 2}
\rho_{1}(0)=\ket{1}\bra{1}=
\left(
\begin{array}{cc}
0 & 0 \\
0 & 1
\end{array}
\right)
\ \Longrightarrow \
\rho_{1}(\infty)=\frac{1}{|O|}
\left(
\begin{array}{cc}
\widehat{O}_{11} & \widehat{O}_{12} \\
\widehat{O}_{13} & \widehat{O}_{14}
\end{array}
\right).
\end{equation}

\vspace{3mm}
As a result
\begin{equation}
\label{eq:final result}
\rho_{0}(0)=\ket{0}\bra{0}, \quad
\rho_{1}(0)=\ket{1}\bra{1}
\ \Longrightarrow \ 
\rho_{0}(\infty)=\rho_{1}(\infty).
\end{equation}

\vspace{3mm}\noindent
Clearly, the Copenhagen interpretation does not hold 
(see the equation (\ref{eq:special last form})). 
We would like to interpret the final density matrix as 
``classical one", see \cite{KF2}.

We

\section{Concluding Remarks}
In this paper we tried to prove the Copenhagen interpretation 
of Quantum Mechanics. In our understanding measurement is 
a kind of decoherence forced and our method is performed 
by embedding it into decoherence theory (which is reasonable 
at least to the author).

We treated the master equation based on density matrix and 
introduced a decoherence time $t_{D}$ (which is in general 
small). Since measurement must be done within $t_{D}$ 
we have only to obtain not the full solution but the approximate 
one of the master equation. 

Our solution gave a proof to the Copenhagen interpretation 
under some assumption. In order to prove the assumption we 
must introduce another theory.

Although our method is not complete it will become a starting 
point to give a complete proof to the Copenhagen interpretation 
in the near future. 
Mathematical physicists with strong mission must prove 
the Copenhagen interpretation at any cost.

%%%%%%%%%%%%%
%References%
%%%%%%%%%%%%%


\begin{thebibliography}{99}
%
\bibitem{KF1}K. Fujii :
\newblock ``Proof" of the Copenhagen Interpretation, 
\newblock arXiv:1304.1591 [quant-ph].
%
\bibitem{KF2}K. Fujii :
\newblock Exact Solution of a Master Equation Applied to the 
Two Level system of an Atom, 
\newblock Int. J. Geom. Methods Mod. Phys, {\bf 11} (2014), 1450085 (18 pages), 
\newblock arXiv:1405.2604 [quant-ph].
%
\bibitem{KF3}K. Fujii :
\newblock Superluminal Group Velocity of Neutrinos : Review, Development 
and Problems, 
\newblock Int. J. Geom. Methods Mod. Phys, {\bf 10} (2013), 1250083 (19 pages), 
\newblock arXiv:1203.6425 [physics].
%
\bibitem{PD}P. Dirac : 
\newblock {\bf The Principles of Quantum Mechanics}, 
\newblock Fourth Edition, Oxford University Press, 1958.
%
\bibitem{HG}H. S. Green : 
\newblock {\bf Matrix Mechanics}, 
\newblock P. Noordhoff Ltd, Groningen, 1965.
%
\bibitem{AP}Asher Peres : 
\newblock {\bf Quantum Theory : Concepts and Methods}, 
\newblock Kluwer Academic Publishers, 1995.
%
\bibitem{AH}Akio Hosoya : 
\newblock {\bf Lectures on Quantum Computation} (in Japanese), 
\newblock SGC Library 4, Saiensu-sha Co., Ltd. Publishers (Tokyo), 1999.
%
\bibitem{WZ}W. H. Zurek :
\newblock Decoherence and the transition from quantum to classical, 
\newblock Physics Today, {\bf 44} (1991), 36-44.
%
\bibitem{WS}W. P. Schleich : 
\newblock {\bf Quantum Optics in Phase Space},
\newblock WILEY--VCH, Berlin, 2001.
%
\bibitem{Lind}G. Lindblad : 
\newblock On the generator of quantum dynamical semigroups, 
\newblock Commun. Math. Phys, {\bf 48} (1976), 119.
%
\bibitem{GKS} V. Gorini, A. Kossakowski and E. C. G. Sudarshan : 
\newblock Completely positive dynamical semigroups of N--level systems, 
\newblock J. Math. Phys, {\bf 17} (1976), 821.
%
\bibitem{BP}H. -P. Breuer and F. Petruccione : 
\newblock {\bf The theory of open quantum systems}, 
\newblock Oxford University Press, New York, 2002.
%
\bibitem{KH}K. Hornberger :
\newblock Introduction to Decoherence Theory, 
\newblock in ``Theoretical Foundations of Quantum Information", 
Lecture Notes in Physics, {\bf 768} (2009), 221-276, Springer, Berlin, 
\newblock quant-ph/061211.
%
\bibitem{CZ}C. Zachos :
\newblock Crib Notes on Campbell-Baker-Hausdorff expansions, 
\newblock unpublished, 1999,
\newblock see\ http://www.hep.anl.gov/czachos/index.html.
%
\bibitem{Five}K. Fujii and et al : 
\newblock {\bf Treasure Box of Mathematical Sciences} (in Japanese),
\newblock Yuseisha, Tokyo, 2010. \\
\newblock I expect that the book will be translated into English.
\end{thebibliography}
\end{document}